\documentclass[letterpaper, 10 pt, conference]{ieeeconf}
\IEEEoverridecommandlockouts
\usepackage{cite}
\usepackage{amsmath}
\usepackage{amssymb}
\usepackage{algorithm}
\usepackage{algpseudocode}
\usepackage{amsmath,amssymb,amsfonts}
\usepackage{mathtools}
\usepackage{graphicx}
\usepackage{textcomp}
\usepackage{xcolor}
\usepackage{tcolorbox}
\usepackage{booktabs}
\usepackage{geometry}
\usepackage{booktabs}   % professional table
\usepackage{multirow}   % multirow cells
\usepackage{makecell}   % optional, for multi
\usepackage{hyperref}
\usepackage[flushleft]{threeparttable}
\def\BibTeX{{\rm B\kern-.05em{\sc i\kern-.025em b}\kern-.08em
    T\kern-.1667em\lower.7ex\hbox{E}\kern-.125emX}}
\begin{document}

\title{\LARGE \bf Robust Restless Multi-Armed Bandit for Data Center Flexibility Services Through Virtual Machine Scheduling\\
}

\author{Zixi Chen$^{*}$,
        Yifu Ding$^{*, \dagger}$,
        Thomas~Magnanti% <-this % stops a space
\thanks{$^{*}$Zixi Chen and Yifu Ding contributed equally to this work. Authors are listed in alphabetical order. $^{\dagger}$Corresponding author: Yifu Ding}
\thanks{Yifu Ding is supported by the MIT Shell Energy Scholar Program at the MIT Energy Initiative.}
\thanks{Zixi Chen is with the School of Mathematics Science, Peking University. Yifu Ding is with MIT Energy Initiative and MIT Sloan School of Management. Thomas Magnanti is with the MIT Sloan School of Management.
e-mail: chenzixi22@stu.pku.edu.cn, yifuding@mit.edu, magnanti@mit.edu.}}
\maketitle

% \begin{IEEEkeywords}
% EV Charging scheduling; Decision-dependent uncertainty; ML classification; Contextual (Stochastic) optimization
% \end{IEEEkeywords}

\begin{abstract}
%% Text of abstract
Energy demands from data centers have surged
and stressed the grid in recent years. Electric grids require balancing supply and demand every second, motivating demand response (reduction) from large loads, including data centers. This can be achieved by rescheduling jobs on a physical machine. Its real-time implementation is uncertain due to
fluctuating resource utilization, and rescheduling incurs quality-of-service (QoS) losses that providers are unwilling to disclose. We propose a restless multi-armed bandit (RMAB) framework, in which the grid operator requests load reductions without access to detailed job-rescheduling procedures. Using open-source virtual machine (VM) datasets, we model job arrivals and rescheduling at each data center as a restless arm in a Markov decision process (MDP) and derive Whittle-index-based policies using the learned transition function via Thompson sampling.
To overcome the weakness of an increasingly long learning process due to an enlarged state space, we use a mixed strategy that includes a global upper confidence bound (UCB) and encodes with trust indices to enhance robustness and accelerate learning. Results show that the proposed mixed-strategy algorithm remains robust across varying state-space sizes and consistently outperforms the pure Thompson-Whittle (TW) algorithm, especially when contextual information is noisy. It also demonstrates superior performance compared to the state-of-the-art EXP4 framework. We provided open-source code to ensure reproducibility. 


\end{abstract}

\section{Introduction}

This letter proposes a contextual RMAB framework to coordinate flexible, privacy-preserving data center demand responses via VM rescheduling. A VM is a software-defined machine that operates within a physical server as an independent computing unit, characterized by heterogeneous features such as CPU or GPU utilization and memory usage. Within a data center, VM rescheduling can improve both service quality and resource efficiency by reallocating workloads to more suitable configurations. Recent demonstrations using workload power-orchestration software show that job rescheduling within a \textit{single} data center can directly respond to grid power-reduction signals \cite{colangelo_ai_2025}. In our framework, the grid operator does not observe the underlying rescheduling process but instead has access only to aggregated, batch-level VM information, such as job flexibility tiers and average power consumption. Moreover, grid operators commonly procure flexibility services from multiple data centers to improve cost-effectiveness and operational performance.

\subsection{Related works}

A RMAB models resource allocation problems in which limited resources must be sequentially assigned to multiple stochastic processes whose states evolve over time under uncertainty. Ref.~\cite{yu_deadline_2016} formulates a stochastic deadline scheduling problem as a RMAB framework, where jobs arrive randomly at a service center and are characterized by stochastic rewards and completion deadlines. A key challenge in RMAB problems is their computational intractability as the state space grows. To address this issue, closed-form, near-optimal index policies have been developed, offering computational tractability together with provable bounds on the performance gap relative to the true optimum \cite{whittle_restless_1988}. This involves proving the RMAB problem is indexable and calculating the Whittle index for each arm. 

The index-based RMAB has also been applied to model demand responses in electric grids, where load states are only partially observable. Ref.~\cite{taylor_index_2014-1} was among the first to introduce an index-based policy for demand responses through thermostatically controlled loads (TCLs). More recently, ref.~\cite{chen_contextual_2024} proposed an online learning algorithm for RMAB-based building demand responses that estimates the index-based policy in the presence of unknown state-transition dynamics, assuming access to perfect global and local contextual information.

The main contributions of this paper are threefold. First, we developed a contextual RMAB framework that enables flexible, privacy-preserving data center services using real-world VM datasets. Instead of disclosing detailed job-level features, the grid operator has access to aggregated, batch-level features. Second, we proposed a learning-based index policy augmented with complementary UCB strategies. This significantly improves performance during the early learning stage, particularly in large state-space settings. Third, the proposed algorithm demonstrates strong robustness across various sensitivity analyses, including changes in problem scale, arm-context noise levels, and policy weights.

\section{Methodology}
\subsection{VM (re)scheduling as a Markov decision process}

Given a set of VMs $\mathcal{V} = \{v_1, v_2, ..., v_m\}$, each has heterogeneous power consumptions $p$ and QoS costs $c$ per core hour. These VM jobs are deployed on $\mathcal{N}$ data centers, and we could model (re)scheduling of each data center as MDP, defined by a tuple $(\mathcal{S}, \mathcal{A}, P, R, \beta)$, where $\mathcal{S}$ is the state space, $\mathcal{A}$ is the action space, $P(s' \mid s, a)$ is the transition kernel, $R(s,a)$ is the one-step reward, and $\beta \in (0,1]$ is the discount factor. The grid could take the action $a\in \mathcal{A}$ to request services from each data center. At time $ t \in \mathcal T$, without the grid request, each data center executes a batch of $N_j$ VM tasks in cyclic job queues consecutively.  If the grid requests the service, the data center responds by looking ahead to the next $N_f \geq N_j$ jobs and selecting the jobs with the lowest power consumption, using the saved power to meet the grid request. The detailed definitions of MDP is given by,

\begin{itemize}
    \item \textbf{States $(s_t)$}: Position $s_i$ in the circular job queue, representing the current batch of $N_j$ jobs.
    \item \textbf{Actions $(a_t)$:} A binary decision variable \(a_i\) indicates whether data center \(i\) receives a request from the grid. When \(a_i = 0\), data center \(i\) executes the default next \(N_j\) jobs in the queue. The data center is activated if \(a_i = 1\) or passive if \(a_i = 0\).
    \item \textbf{Rewards $(r_t)$:} When action \(a_i = 0\), the reward is always zero. When action \(a_i = 1\), the reward is defined as the net benefit from rescheduling, which is computed as the power savings minus the delay penalty associated with any interactive jobs skipped during reordering. This is detailed in Appendix, Section \ref{tab:rescheduling}.
\end{itemize}

A few underlying assumptions are made: (1) the state space of data centers is finite, implying that the job queue at each data center follows a recurring pattern over time; (2) rescheduling under the same hardware configuration does not alter power consumption of VM jobs; and (3) the framework models real-time interactions between the grid operator and data centers, where each RMAB learning round corresponds to a grid request. We assume that the duration of each demand response event is shorter than the interval between consecutive requests, such that individual grid requests can be treated as independent.

\subsection{Model formulation}

\subsubsection{RMAB problem}
At time $t \in \mathcal{T}$, the grid sends requests to a pre-determined, limited number of data centers $N_t \leq \mathcal{N}$, depending on the grid conditions. The RMAB problem $\mathcal M$ is formulated as, 

\begin{equation} 
\begin{aligned}
(\mathcal M) \quad \text{obj.}  & \quad \max_{\pi}\mathbb{E}_{\pi}  \sum^{\infty}_{t=0} \sum^{\mathcal{N}}_{i=1}   \{\beta^t r_{i, t} a_{i, t}\} \\
 \text{s.t.} &  \quad   \sum^{\mathcal{N}}_{i=1}  a_{i, t} \leq N_{t}\\
& \quad r_{i, t} = R_i (a_{i,t}, s_{i,t})
\label{objective3}
\end{aligned}
\end{equation}

The objective is to design a control policy $\pi$ that maximizes the long-term cumulative discounted reward, given the per-round reward $r_{i,t}$ and discount factor $\beta$. The state transition of each arm depends on whether the corresponding data center is activated. Specifically, $P_i^1(\cdot \mid s)$ denotes the transition function under activation, while $P_i^0(\cdot \mid s)$ denotes the transition function when the arm remains passive.
% \begin{equation}
% s_{i,t+1}\sim
% \begin{cases}
% P_i^1(\cdot\mid s_{i,t}), & y_{i,t}=1,\\
% P_i^0(\cdot\mid s_{i,t}), & y_{i,t}=0.
% \end{cases}
% \end{equation}

In our framework, the reward function $R_i$ and transition function are assumed to be unknown; otherwise, the problem reduces to a standard dynamic optimization problem.

\textit{Lamma 1 (Intractability):} In the optimization problem $\mathcal M$, if each data center has its own state space $\mathcal{S}$ and actions affect future states, then selecting $n$ data centers per round leads to a joint state space that grows as $|\mathcal{S}|^n$. As a result, solving the full optimization quickly becomes intractable.

\subsubsection{The curse of dimensionality and Whittle index}

% If each data center has its own state space $\mathcal{S}$ and actions affect future states, then selecting $n$ data centers per round leads to a joint state space that grows as $|\mathcal{S}|^n$. As a result, solving the full optimization quickly becomes intractable.

% Whittle's index:

Considering the curse of dimensionality in the problem $\mathcal M$, Whittle \cite{whittle_restless_1988} proposed an index policy to strategically calculate the marginal reward for activating each arm independently based on Lagrangian relaxation, which is as follows, 

\textit{Definition 1 (Whittle index \cite{whittle_restless_1988})}: 
For each arm $i$ at state $s$, consider a single-arm MDP with a passivity
subsidy $\lambda$. Let $V_{i,\lambda}(s)$ be the optimal discounted value.
The Whittle index $W_i(s)$ is the smallest subsidy that makes passivity
optimal:
\begin{align}
\label{eq:subsidy_mdp}
W_i(s) = \inf \Bigg\{ \lambda :\;
&\lambda + \beta \sum_{s'} P_i^0(s,s') V_{i,\lambda}(s') \nonumber \\
&\ge r_i(s) + \beta \sum_{s'} P_i^1(s,s') V_{i,\lambda}(s')
\Bigg\}.
\end{align}
Under indexability, arms are ranked by $W_i(s)$ and the highest-index arms are
activated each round. Equivalently, define
\begin{align}
Q_{i,\lambda}^{1}(s)
&=
R_i(s,1)
+\beta \sum_{s'} P_i^{1}(s,s')V_{i,\lambda}(s'),
\\
Q_{i,\lambda}^{0}(s)
&=
R_i(s,0)+\lambda
+\beta \sum_{s'} P_i^{0}(s,s')V_{i,\lambda}(s').
\end{align}
For a given $\lambda$, the passive-optimal set is
\begin{equation}
\mathcal P_i(\lambda)
=
\{s\in\mathcal S_i: Q_{i,\lambda}^{0}(s)\ge Q_{i,\lambda}^{1}(s)\}.
\end{equation}

Arm $i$ is \emph{indexable} if $\mathcal P_i(\lambda)$ is nondecreasing in
$\lambda$ by set inclusion and expands from $\emptyset$ to $\mathcal S_i$ as
$\lambda$ increases from $-\infty$ to $+\infty$. Thus, for any
$\lambda_1\le \lambda_2$,
\[
\mathcal P_i(\lambda_1)\subseteq \mathcal P_i(\lambda_2).
\]
Under indexability, the Whittle index can also be written as
\begin{equation}
W_i(s)
=
\inf\{\lambda:\; Q_{i,\lambda}^{0}(s)\ge Q_{i,\lambda}^{1}(s)\}.
\label{eq:whittle_index}
\end{equation}
At each round $t$, the aggregator activates the $N_t$ arms with the largest
indices $W_i(s_{i,t})$.

\textit{Proposition 1 (Sufficient condition for indexability):}
For arm $i$, define
\begin{equation}
\Delta_{i,\lambda}(s)
=
Q_{i,\lambda}^{1}(s)-Q_{i,\lambda}^{0}(s),
\qquad s\in\mathcal S_i .
\end{equation}
If $\Delta_{i,\lambda}(s)$ is continuous and non-increasing in $\lambda$ for
every $s\in\mathcal S_i$, and there exist $\lambda_{\min}$ and
$\lambda_{\max}$ such that all states are active-optimal for
$\lambda\le \lambda_{\min}$ and passive-optimal for
$\lambda\ge \lambda_{\max}$, then arm $i$ is indexable.

\textit{Proof:}
Since $\Delta_{i,\lambda}(s)$ is continuous and non-increasing in $\lambda$,
each state switches from active-optimal to passive-optimal at most once. Hence
\[
\mathcal P_i(\lambda)
=
\{s\in\mathcal S_i:\Delta_{i,\lambda}(s)\le 0\}
\]
expands monotonically with $\lambda$. The boundary conditions imply that
$\mathcal P_i(\lambda)$ evolves from $\emptyset$ to $\mathcal S_i$, so arm $i$
is indexable.

In our implementation, indexability is checked numerically by solving
\eqref{eq:subsidy_mdp} over a subsidy grid and verifying that
$\mathcal P_i(\lambda)$ expands monotonically.

%\paragraph{Contextual bandit}

% We construct a four-dimensional context vector for each decision:
% \begin{itemize}
%     \item \textbf{Average power:} mean of \textit{avgcpu} $\times$ \textit{corehour\_norm} for the current batch of five jobs, capturing how power-intensive the batch is.
%     \item \textbf{Average normalized core-hour:} mean normalized core-hour for the current batch, serving as a proxy for expected runtime.
%     \item \textbf{Fraction of interactive jobs:} share of Interactive VM jobs in the batch (0 to 1), indicating the potential delay penalty from reordering.
%     \item \textbf{Normalized state:} $s/\textit{num\_jobs}$, the queue position scaled to $[0,1]$, capturing position-dependent non-stationarity.
% \end{itemize}
\subsubsection{Strategies for arm selection}

\paragraph{Contextual bandit}
As a model-free baseline, we consider a contextual bandit that uses only the
current batch features and ignores state transition functions. For each arm $i$ at time $t$, we define the context vector $x_i(t) \in \mathbb{R}^n$ of each VM job batch. We model the expected reward of each arm as a \textit{linear} function of a context (feature) vector and maintain a Bayesian posterior over $\theta_i$.
% \begin{equation}
% x_i(t)
% =
% \big[
% \bar p_i(t),\;
% \bar h_i(t),\;
% d_i(t),\;
% s_i(t)/N_s
% \big]^\top,
% \end{equation}
% where $\bar p_i(t)$ is the average power consumption of the current batch,
% $\bar h_i(t)$ is the average normalized core-hour, $d_i(t)$ is the fraction of
% delay-sensitive jobs, and $s_i(t)/N_s$ is the normalized queue position.

\begin{equation}
r_{i,t}=x_i(t)^\top \theta_i+\varepsilon_{i,t},
\qquad
\varepsilon_{i,t}\sim\mathcal N(0,\sigma^2),
\end{equation}
where $\varepsilon_{i,t}$ represents the observation noise atop the context information. At time $t$, we sample
$\tilde\theta_i(t)$ from the posterior and compute the sampled score,
\begin{equation}
\tilde\mu_i(t)=x_i(t)^\top\tilde\theta_i(t).
\end{equation}
 The contextual Thompson policy then activates the $N_t$ arms with the highest
sampled scores $\tilde\mu_i(t)$. 

\paragraph{Thompson--Whittle (TW)}

For each arm $i$ at time $t$, a corresponding state $s_i(t) := \psi(x_i(t)) \in\mathcal S_i$ is observed from the context mapping $\psi(x_i(t))$. The arm model is represented by unknown reward and transition functions $\theta_i=(r_i,P_i^1,P_i^0)$. The algorithm first observes current states through the contextual vector of each arm and samples from the posterior distribution to obtain estimates of the reward and transition functions. The transition dynamics are modeled using a Dirichlet--Categorical prior, while the reward function is represented by a Gaussian prior. At each learning step, one sample is drawn from the posterior to compute the corresponding Whittle index. Weakly informative priors are used to enable smoothing in sparsely visited states while remaining adaptive to newly observed data. For each arm \(i\) and subsidy \(\lambda\), the sampled single-arm subsidy MDP is solved.

\begin{align}
\tilde V_{i,\lambda}^{(t)}(s)
= \max \Big\{
&\tilde r_i(s)
+ \beta\sum_{s'} \tilde P_i^1(s,s')
  \tilde V_{i,\lambda}^{(t)}(s'), \nonumber\\
&\lambda
+ \beta \sum_{s'} \tilde P_i^0(s,s')
  \tilde V_{i,\lambda}^{(t)}(s')
\Big\}.
\label{MDP_subsidy}
\end{align}

Then, the Whittle index is computed as (\ref{eq:subsidy_mdp}).

\begin{algorithm}[!h]
\caption{TM-TW Index Policy}
\label{alg:tmtw}
\begin{algorithmic}[1]

\State Initialize prior $\pi_1(\theta)$; Delay horizons $T_{\epsilon}^{\mathrm{mix}}$ and $T_g$
\State Set index update period $N_{\text{up}}$

\For{$t = 1,2,\dots,T$}

\State Observe current states $s_i(t)$ for all arms

\For{each arm $i$}

\State Sample posterior using historical information
\begin{equation}
\tilde{\theta}_t
=
\{(\tilde r_i,\tilde P_i^1,\tilde P_i^0)\}_{i=1}^N
\sim
\pi_t(\cdot \mid \mathcal H_t) \nonumber
\label{posterior}
\end{equation}

\If{$t \bmod N_{\text{up}} = 1$}

\State Solve the sampled subsidy MDP (\ref{MDP_subsidy}) 
% \begin{align}
% \tilde V_{i,\lambda}^{(t)}(s)
% =&
% \max
% \Big\{
% \tilde r_i(s)
% + \beta \sum_{s'} \tilde P_i^1(s,s')
% \tilde V_{i,\lambda}^{(t)}(s'), \nonumber\\
% &\lambda
% + \beta \sum_{s'} \tilde P_i^0(s,s')
% \tilde V_{i,\lambda}^{(t)}(s')
% \Big\}
% \end{align} 
\State Compute the Whittle index $\tilde W_i (t)$
\State Compute the mixed score $S(t)$.
\EndIf

\EndFor

\State Rank all arms based on normalized mixed scores.
\State Activate the top $N_t$ arms by the ranking order.

\State observe $(R_t,S(t+1))$,  update posterior $\pi_{t+1}$
% \begin{align*}
% \pi_{t+1}(\theta \mid \mathcal H_{t+1})
% &\propto p\!\left(R_t, S(t+1) \mid S(t), A(t), \theta\right) \times \pi_t(\theta \mid \mathcal H_t)
% \end{align*}

\EndFor

\end{algorithmic}
\end{algorithm}

\paragraph{Trust-Mixed Thompson--Whittle (TM--TW)}
% To accelerate early exploration while preserving Whittle-structured exploitation, we define a trust-controlled convex mixture between normalized sampled Whittle and greedy scores:
% \[
% \hat S_i(t)=\tau_t\,\hat W_i(t)+(1-\tau_t)\,\hat G_i(t),\qquad
% i_t\in\arg\max_i \hat S_i(t),
% \]
% where $\hat W_i(t),\hat G_i(t)\in[0,1]$ are min--max normalized scores across arms at round $t$.

% The greedy score combines a global arm-quality UCB and a local state-wise UCB:
% \[
% G_i(t)=w_t\!\left(\bar\mu_i(t)+c_g\sqrt{\frac{\log(t+2)}{n_i(t)+n_0}}\right)
% +(1-w_t)\!\left(\mu_{i,s_i(t)}+c_\ell \sigma_{i,s_i(t)}\right),
% \]
% with decay
% \[
% w_t=w_0\max\!\left\{0,1-\frac{t}{T_g}\right\}.
% \]
% Hence TM--TW is more greedy early (large $w_t$), then gradually shifts to local posterior structure.

% The trust index $\tau_t$ increases with learning progress, posterior confidence, and Whittle-table stability:
% \[
% \tau_t=\mathrm{clip}_{[\tau_{\min},\tau_{\max}]}
% \left(
% \max\!\left\{
% (0.25+0.75\rho_t)\big(\omega_c C_t+\omega_p P_t+\omega_s S_t\big),\;
% \tau_{\max}\,\sigma\!\left(\frac{t-t_0}{T_0}\right)
% \right\}
% \right),
% \]
% where $C_t$ is model confidence (reward and transition observations), $P_t$ is normalized round progress, $S_t=\exp(-\Delta W_t/\kappa_w)$ measures Whittle stability, and $\rho_t$ is active-coverage confidence. Therefore TM--TW explicitly implements ``explore fast first, then trust Whittle more''.

%One could observed that compared to the model-free contextual bandit, the Thompson-Whittle could outperform by introducing an index policy atop the learned transition functions.

TW policy has a practical weakness: in the early rounds,
the sampled transition model can be noisy, which makes the estimated Whittle
indices volatile. We address
this issue by blending the sampled Whittle score with a greedy uncertainty-aware
score and gradually increasing the weight on the Whittle component as evidence
accumulates. The intent is to exploit readily observable reward signals at the
start of learning, without discarding the long-horizon value of the RMAB model. 

First, we define a linear combination of the normalized Whittle and greedy scores, which represents the trust in two types of arm-selection strategies.
\begin{align}
\hat S_i(t)
= (1-\tau_t)\hat W_i(t)
+ \tau_t \hat G_i(t),
\end{align}

% \begin{align}
% i_t \in \arg\max_i \hat S_i(t),
% \end{align}

Where $\hat W_i(t),\hat G_i(t)\in[0,1]$ are min--max normalized scores. The mixing coefficient $\tau_t$ controls the transition from the greedy-UCB
score to the TW score, and decays linearly until the policy becomes the pure TW:
\begin{align}
\tau_t
= \textstyle
\max\!\left\{
0,\,
1-\frac{t}{T_{\epsilon}^{\mathrm{mix}}}
\right\}.
\end{align}

The greedy score $G_i(t)$ combines the global and local contextual UCB estimates. These two algorithms select arms based on the expected reward derived from the entire historical information and the current local state, respectively.
\begin{align}
G_i(t)
= 
&  \textstyle \underbrace{w_t\!\left(
\bar\mu_i(t)
+
c_g\sqrt{\frac{\log(t+2)}{n_i(t)+n_0}}
\right)}_{\text{global UCB}}
\nonumber\\
+
&\underbrace{(1-w_t)\!\left(
\mu_{i,s_i(t)}
+
c_\ell \sigma_{i,s_i(t)}
\right)}_{\text{local UCB}}.
\end{align}

The weight $w_t$ controls the transition within the greedy score from the global to local exploration:
% \begin{align}
% w_t
% =
% \max\!\left\{
% 0,\,
% 1-\frac{t}{T_g}
% \right\}.
% \end{align}
\begin{align}
w_t = \textstyle \max\!\left\{0,\, 1-\frac{t}{T_g}\right\}
\end{align}

Thus, the period parameter $T_g$ governs the shift from the global UCB term to the local UCB term
inside the greedy score, while $T_{\epsilon}^{\mathrm{mix}}$ governs the shift
from greedy UCB to TW. In particular, when $t \ge T_{\epsilon}^{\mathrm{mix}}$,
we have $\tau_t=0$, so $\hat S_i(t)=\hat W_i(t)$ and the policy reduces to
pure TW. Finally, TM--TW activates the $N_t$ arms with the highest mixed scores
$\hat S_i(t)$. The full procedure is summarized in Algorithm \ref{alg:tmtw}. In this way, the policy first exploits coarse global information, then gradually emphasizes local information, and eventually
transitions to the learned Whittle structure.

\section{Case study}

% \begin{itemize}
%     \item \textit{Average power consumption}: mean of power consumption for the current batch of $N_j$ jobs, capturing how power-intensive the batch is.
%     \item \textit{Average normalized core-hour:} mean of the normalized core-hour for the current batch, serving as a proxy for expected power consumption and reward if activated.
%     \item \textit{Fraction of delay-sensitive jobs:} share of delay-sensitive jobs in the current batch, indicating the potential delay penalty from reordering.
%     \item \textit{Normalized state:} $s/\textit{num\_jobs}$, the queue position scaled to $[0,1]$, capturing position-dependent non-stationarity.
% \end{itemize}

\begin{figure}[!h]
    \centering
    \includegraphics[width=\linewidth]{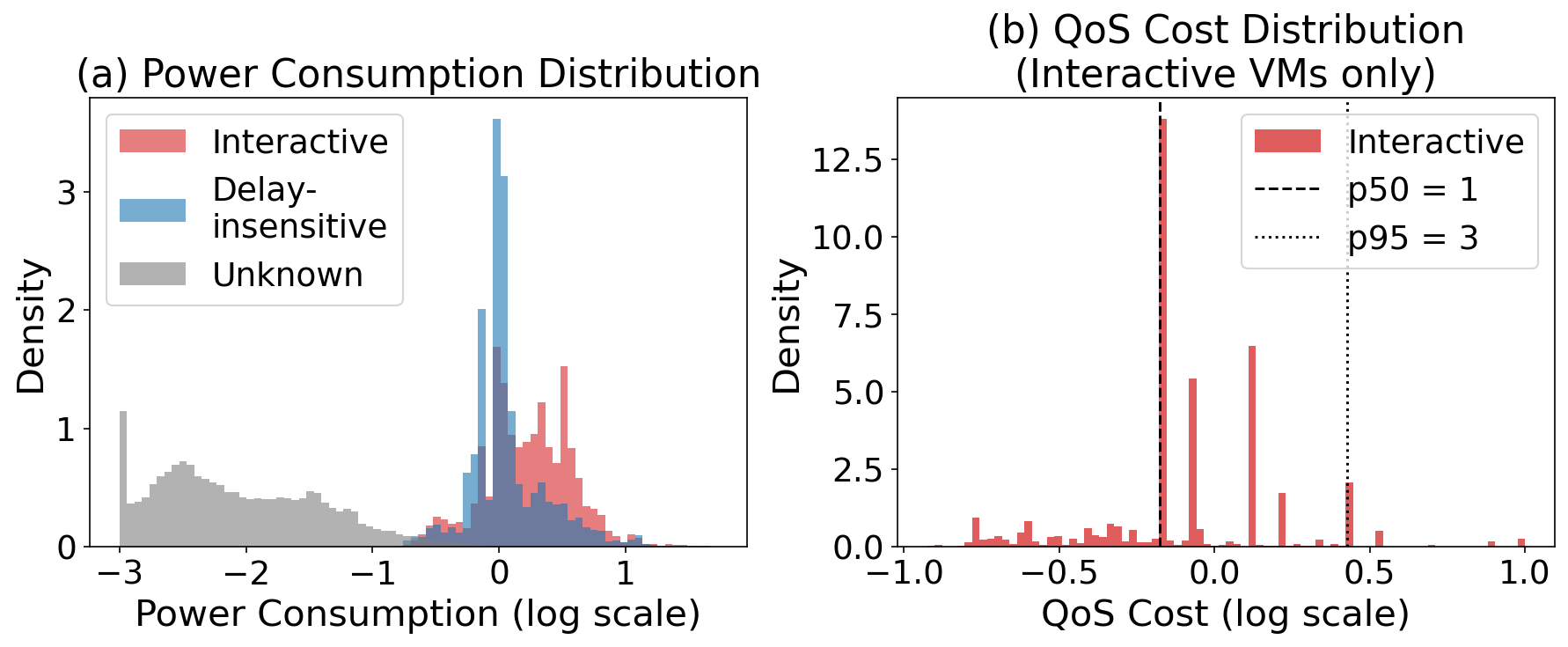}
    \caption{Distributions of (a) normalized power consumption and (b) QoS cost across VM traces. Power consumption is computed from CPU utilization and normalized core-hour usage, while QoS cost is evaluated for Interactive VMs only. Both are shown on a log scale.}
    \label{fig:power_and_qos}
\end{figure}

We use the Microsoft Azure VM Dataset reported in \cite{cortez_resource_2017} for the case study. VM traces with less than 1 core-hour of usage and utilization below 10\% have been filtered out. For each VM job, we estimate its power consumption and QoS costs. Fig. \ref{fig:power_and_qos}  (a) and (b) show distributions of power consumption and QoS cost of VM traces. Power consumption per core-hour is modeled as a piecewise linear function of CPU utilization. The effects of data center location and power network constraints are incorporated into the reward function via a proxy locational electricity price, $\lambda_{\mathrm{LMP}}$. The QoS cost is calculated as the product of the job-specific per-core-hour QoS cost, $q_j$, and the corresponding core-hour consumption of each VM job, as detailed in Appendix~\ref{tab:power_consumption}.

The case study models $N_m$ VM jobs execution as a batch, and this gives each data center $N_s := \frac{N_m}{N_j}$ independent states. We assume that the following local contextual information is accessed for each batch: (a) average power consumption, (b) average normalized core-hour, (c) fraction of delay-sensitive jobs, and (d) normalized state scaling between 0 and 1, which shows the position in the job queue. 

To enhance reproducibility, we provided an open-source code for our framework \textit{RACER}, a \textbf{\underline{r}}obust and \textbf{\underline{a}}daptive \textbf{\underline{c}}omputing and \textbf{\underline{e}}nergy \textbf{\underline{r}}esource coordination framework: \url{https://github.com/Yifueveding/RACER} / \url{https://zenodo.org/records/20543532}.

We first varied the number of data centers and activation budgets across experiments. Specifically, the system size was set to 3, 5, 8, and 10 data centers. The activation budget scaled proportionally to system size, allowing selection of 1 out of 3, 2 out of 5, 3 out of 8, and 4 out of 10 data centers per round, respectively. For each random seed and system-size configuration, we compared the proposed TM--TW strategy against three benchmark methods: (a) Oracle Whittle (Oracle), which assumes the perfect knowledge of the true transition and reward functions and serves as an upper-bound benchmark; (b) TW, which learns the transition model and computes Whittle indices from posterior samples using contextual information; and (c) ST, which directly learns state-wise rewards from context vectors without exploiting the index structure.

\begin{table}[htbp]
\centering
\caption{Reward comparison of TM--TW, TW, and ST with 40 sampled VM jobs per data center}
\label{tab:local_global_tw_n40}
\setlength{\tabcolsep}{4pt}
\small
\resizebox{\linewidth}{!}{\begin{tabular}{cc ccc ccc}
\toprule
\multirow{2}{*}{$n_{\mathrm{dc}}$} &
\multirow{2}{*}{$N_t$} &
\multicolumn{2}{c}{TW} &
\multicolumn{2}{c}{ST} &
\multicolumn{2}{c}{TM--TW} \\
\cmidrule(lr){3-4}
\cmidrule(lr){5-6}
\cmidrule(lr){7-8}
&
& Reward & Oracle (\%)
& Reward & Oracle (\%)
& Reward & Oracle (\%) \\
\midrule
3  & 1 & 6.619 & 89.37 & 5.765 & 77.85 & 6.633 & 89.57 \\
5  & 2 & 28.960 & 97.84 & 28.419 & 96.01 & 29.010 & 98.00 \\
8  & 3 & 30.397 & 92.48 & 28.705 & 87.33 & 30.674 & 93.32 \\
10 & 4 & 40.353 & 94.27 & 37.465 & 87.52 & 41.267 & 96.41 \\
\bottomrule
\end{tabular}}
\end{table}

\begin{figure}[!h]
    \centering
    \includegraphics[width=\linewidth]{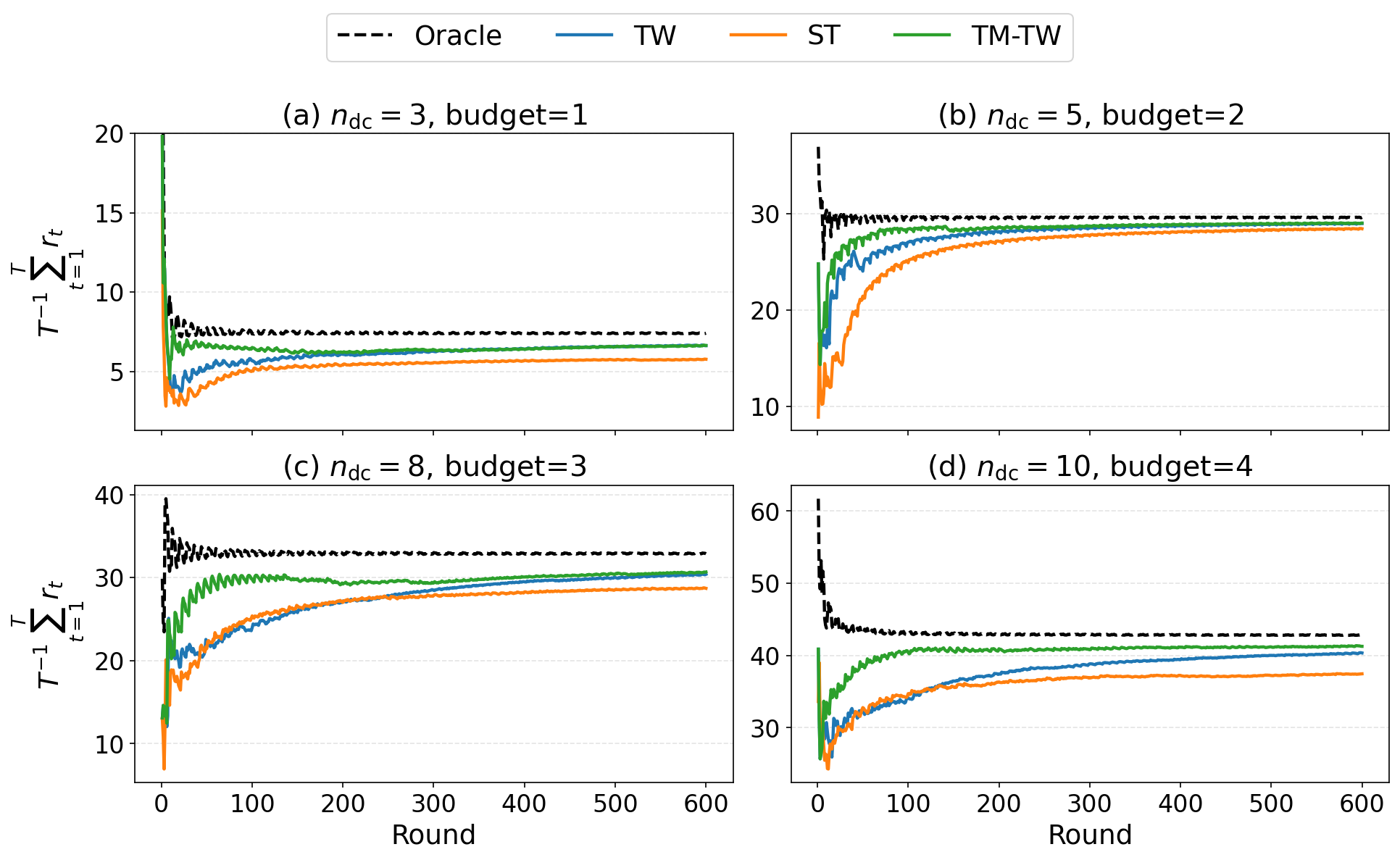}
    \caption{Running average reward over 600 rounds for Oracle, TW, ST, and TM--TW under four data center--budget configurations: (a) \(n_{\mathrm{dc}} = 3\), budget = 1; (b) \(n_{\mathrm{dc}} = 5\), budget = 2; (c) \(n_{\mathrm{dc}} = 8\), budget = 3; and (d) \(n_{\mathrm{dc}} = 10\), budget = 4. Each curve represents the reward averaged across two random seeds with 40 sampled VM jobs per data center.}
    \label{fig:tw_vs_blackbox_high}
\end{figure}

Table~\ref{tab:local_global_tw_n40} summarizes the performance comparison among three learning strategies: TM--TW, TW, and ST. Across all four activation budget  configurations, TM--TW consistently achieves the highest rewards, followed by TW, while ST performs the worst in every setting. Relative to TW, TM--TW improves performances by \(0.19\%\), \(0.17\%\), \(0.85\%\), and \(2.14\%\) across the four configurations, respectively. Fig.~\ref{fig:tw_vs_blackbox_high} illustrates the running average rewards of three strategies relative to the Oracle over 600 rounds. TW initially performs below the Oracle policy during the learning phase, but its reward trajectory steadily improves as the posterior uncertainty is reduced. In contrast, ST rapidly converges to lower-reward regimes and exhibits persistently higher regret. In all settings, TM--TW consistently outperforms TW by substantially improving learning performance in the early stage.

\subsection{Ablation experiments on the proposed TM--TW algorithm}

We next conducted ablation experiments on the proposed TM--TW algorithm. The Oracle Whittle policy serves as the benchmark, and performance is evaluated against a range of baseline algorithms, including ST and TM--TW variants with global and local UCB components. We also compare the proposed TM--TW strategy, which relies on decayed trust weights, with the state-of-the-art mixed-strategy framework, the Exponential-weight Algorithm for Exploration and Exploitation using Expert advice (EXP4)~\cite{beygelzimer_contextual_2011}. EXP4 assumes a collection of experts, each of which provides a recommendation or a probability distribution over arms at every round. The learner then adaptively updates the experts' weights based on their observed performance. In our setting, the Global UCB, Local UCB, and TW policies are treated as three experts within the EXP4 framework, each offering arm-selection recommendations based on its own decision logic. EXP4 dynamically updates the weights assigned to these experts over time and aggregates their recommendations to determine the final arm-selection policy.

\begin{table}[htbp]
\centering
\caption{Performance comparison relative to the Oracle Whittle benchmark (ranked by average reward).}
\label{tab:policy_comparison_2}
\setlength{\tabcolsep}{6pt}
\resizebox{\linewidth}{!}{\begin{tabular}{lccc}
\toprule
  Policy & \makecell{Avg. reward \\ (\$ /round)} & \makecell{Cum. reward / \\ Oracle (\%) }& Avg. sim time (s) \\
  \midrule
Oracle Whittle      & 16.139 & 100.00 & 0.039 \\
Local + Global + TW & 15.465 & 95.82  & 100.904 \\
Global UCB + TW     & 15.360 & 95.17  & 100.966 \\
Local UCB + TW      & 14.687 & 90.97  & 102.378 \\
EXP4-Strategy \cite{beygelzimer_contextual_2011} & 12.814 & 79.40  & 104.697 \\
TW                  & 15.276 & 94.65  & 101.969 \\
ST      & 8.334  & 51.64  & 0.047 \\
\bottomrule
\end{tabular}}
\end{table}

Table \ref{tab:policy_comparison_2} reports an ablation study of the mixed strategies for RMAB problems. The TW policy achieves 94.65\% of the Oracle cumulative reward. Incorporating Global UCB improves it to 95.17\%, suggesting that global exploration provides more effective learning across arms. The combined strategy (Local + Global + TW) achieves the best non-oracle performance, reaching 95.82\% of the Oracle reward. The average simulation time is approximately 100 seconds for all non-oracle index policies, indicating similar computational effort across index-based strategies at a given state-space size. We also tested two decay-horizon parameters in the mixed strategies, $T_g$ and $T_{\epsilon}^{\mathrm{mix}}$, by increasing their magnitudes to 200\%, and the reward changed by 2\%.

\begin{figure}[!h]
    \centering
    \includegraphics[width=\linewidth]{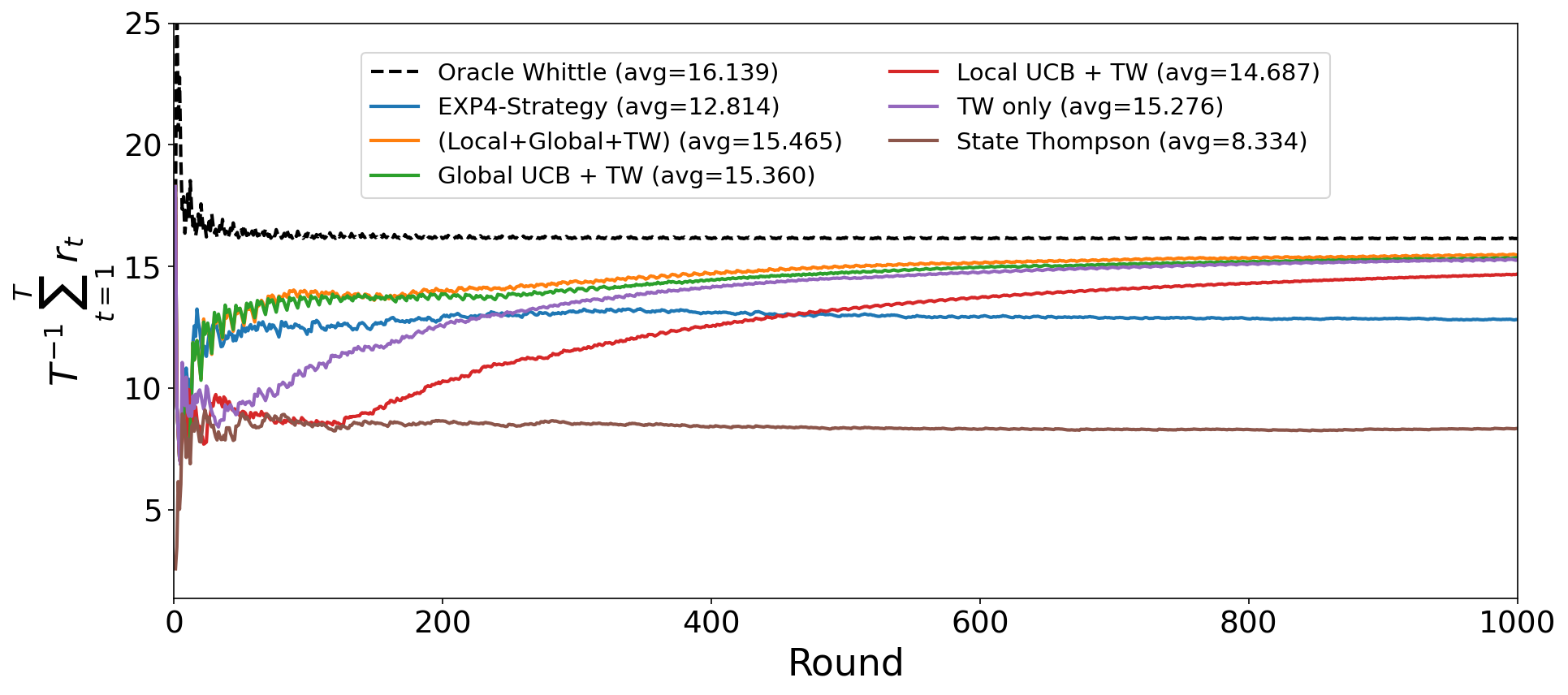}
    \caption{Running average reward over 1000 rounds for EXP4, State Thompson, and Thompson-Whittle ablations. Oracle Whittle is shown as an upper-bound reference. The combined Local+Global+TW variant achieves the strongest learned performance and approaches the oracle most closely.}
    \label{fig:tm_tw_vs_blackbox}
\end{figure}

Fig. ~\ref{fig:tm_tw_vs_blackbox} illustrates that the benefit of trust mixture is primarily concentrated in the transient learning stage. The mixed strategy (i.e., Local + Global + TW) improves substantially faster than TW during the first 100--200 rounds and subsequently stabilizes near the oracle trajectory. Although TW eventually learns the correct arm ranking, its slower initial learning induces a regret gap that is not recovered later. This pattern aligns with the policy design: the greedy component provides a strong short-term signal when the posterior model remains under-trained, while the trust schedule progressively shifts control back to the Whittle index as the learned model becomes more reliable.

We also find that EXP4 performs competitively during the first 100 rounds, as it rapidly identifies and follows the best-performing policy. However, its performance plateaus and is eventually surpassed by the proposed approach. This occurs because EXP4 treats component policies as black-box experts and updates their weights solely based on realized rewards, without leveraging the underlying RMAB structure or transition dynamics. Consequently, EXP4 may adapt more slowly and exhibit higher variance under noisy contextual observations and sparse feedback.

\subsection{Sensitivity analysis}

We next examine whether the performance gains of TM–TW remain robust under changes in the problem scale and policy calibration. Specifically, we consider two practically relevant perturbations: (a) varying the size of the state space by increasing the number of VM jobs involved in the data center scheduling, and (b) introducing different levels of noise to the contextual vectors to contaminate state observations, simulating real-world sensor inaccuracies or data transmission errors. 

\subsubsection{Size of state space}
\begin{figure}[!h]
    \centering
    \includegraphics[width=\linewidth]{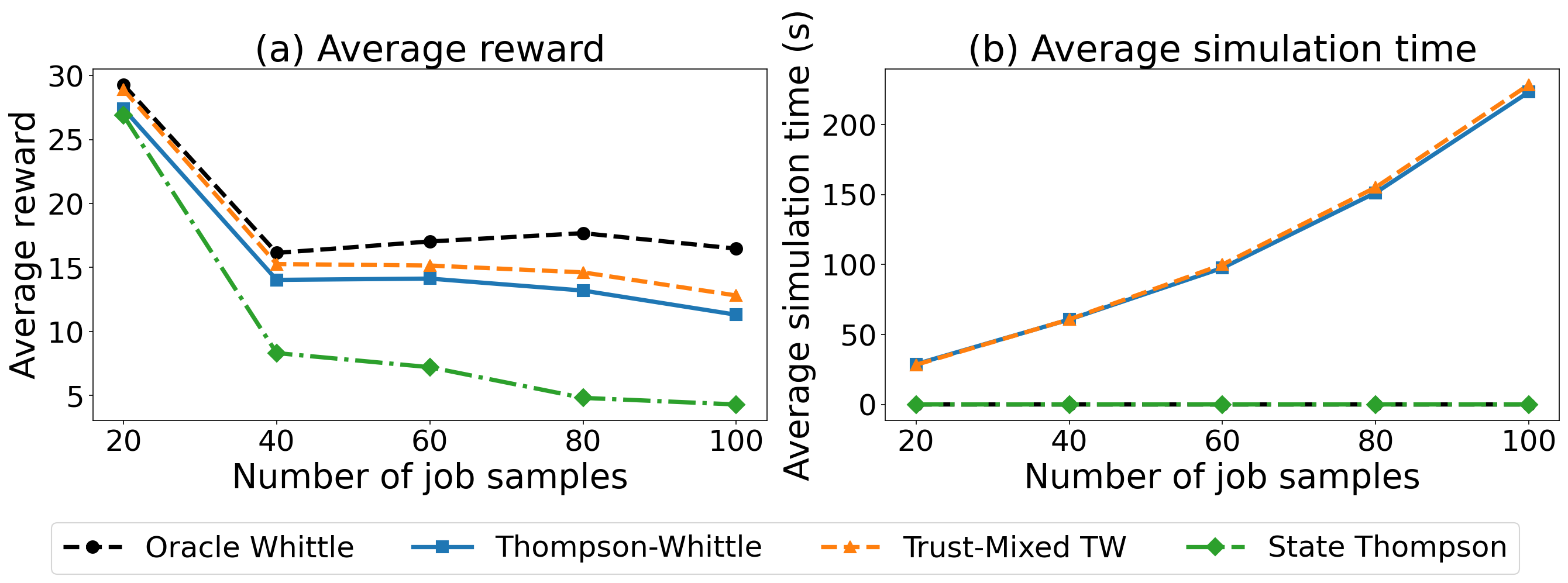}
    \caption{Sensitivity analysis results: (a) the average reward and (b) the average simulation time regarding the number of job samples (state space size) of four strategies: Oracle, TW, TM-TW, and ST.}
    \label{fig:sensitivity_jobs}
\end{figure}

Fig. \ref{fig:sensitivity_jobs} shows the sensitivity analysis results as the number of VM jobs scheduled per cycle increases from 20 to 100. First, the average reward decreases due to the degradation of Thompson learning in the larger state spaces. Nevertheless, the TM-TW policy continues to outperform ST and TM across all settings. Secondly, the required computation time increases in a nearly quadratic trend for the TW and TM-TW policies. This is expected, since each index update requires solving the single-arm subsidy MDP, which has complexity $\mathcal{O}(n^2)$. In contrast, the Oracle requires less than 10 seconds, because the index is computed only once and used for simple lookups in later rounds.

\subsubsection{Level of state observation noise}
\begin{figure}[!h]
    \centering
    \includegraphics[width=\linewidth]{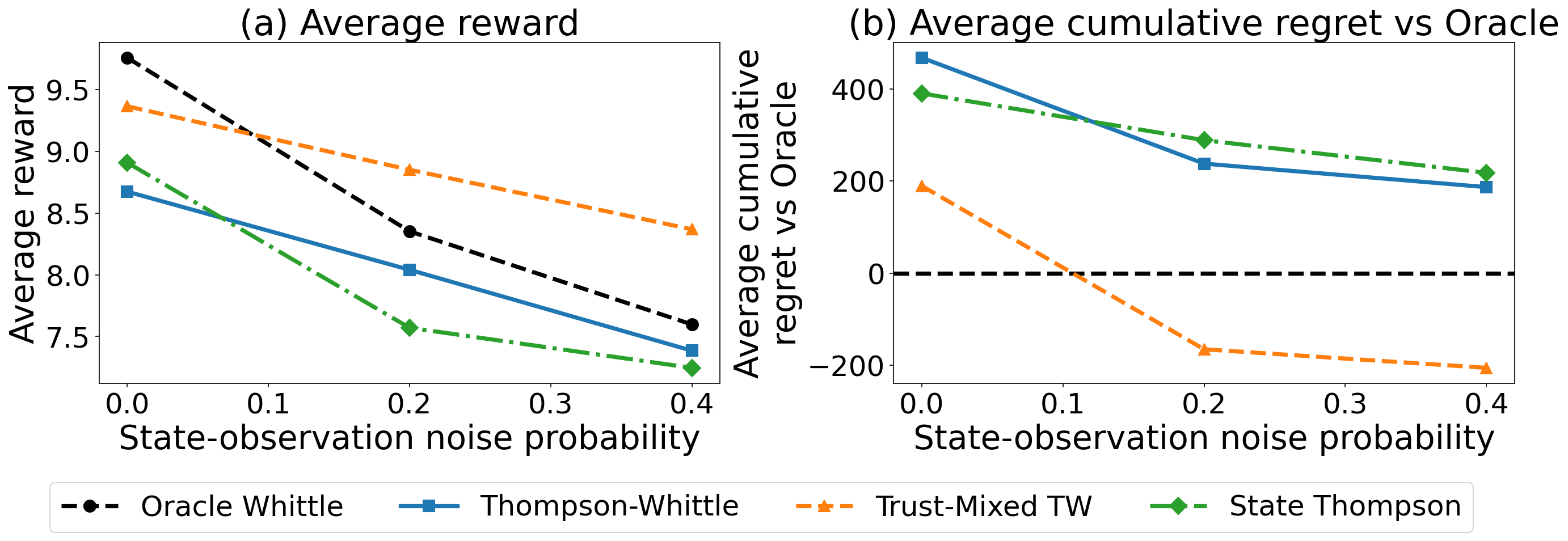}
    \caption{Sensitivity analysis results: (a) the average reward and (b) the average cumulative regret versus oracle regarding the varying state-observation noise level of four strategies: Oracle, TW, TM-TW, and ST}
    \label{fig:sensitivity_noise}
\end{figure}

 We introduce Gaussian noise $\varepsilon_{i,t}$ into the contextual mapping $\psi(x_i)$, so that the policy identifies incorrect states with probability ranging from 0 to 0.4. Fig. \ref{fig:sensitivity_noise} shows that all four policies experience performance degradation. Even Oracle Whittle, which has full knowledge of the true reward function and transition dynamics, suffers a noticeable performance decline, with the average reward per round decreasing from 9.76 to 7.60. This confirms that corrupted state observations directly affect the Whittle index lookup, thereby degrading decision quality.

Among the four policies, TM-TW demonstrates the strongest robustness. Its average reward declines moderately from 9.37 to 8.37 when the noise level reaches 0.4, and it even surpasses Oracle Whittle when the noise level exceeds 0.15. This robustness arises from the greedy UCB component, which combines global arm-pull statistics with local posterior estimates instead of relying solely on state-indexed Whittle scores, thereby mitigating the impact of corrupted contextual information. In contrast, ST has a reduction in the average reward from 8.91 to 7.25, while TW experiences a reduction in the average reward from 8.67 to 7.38.

\section{Discussion}

This paper develops an RMAB framework for data center demand responses under information privacy constraints, where the grid operator observes only batch-level, potentially noisy VM information. We propose a TW index policy that incorporates Thompson learning to estimate the underlying state transitions. To address limited observations during the early learning stage, we further augment the policy with a global--local UCB exploration mechanism through trust-mixed indices. Numerical results demonstrate that the proposed algorithm is robust to varying state-space sizes and noisy contextual information. The TM--TW algorithm consistently outperforms the classical EXP4 strategy, which reweights strategies solely based on historical performance in a black-box manner. However, the performance improvement of TM--TW over the original TM policy varies across settings, suggesting the need for further regret analysis to better understand its theoretical guarantees, which will be our future work.

  %\begin{thebibliography}
\bibliographystyle{IEEEtran}
\bibliography{CDC_data_center_bandit}
%\end{thebibliography}

\section{Appendix}

\subsection{Power consumption of VM workloads}
\label{tab:power_consumption}

The power consumption of each VM job per core-hour is modeled as a piecewise-linear function of CPU or GPU utilization, comprising a static component and a utilization-dependent dynamic component.
\[
    P_{tot}
    = P_{static} + \frac{P_{max}- P_{static}}{u_{max}- u_{min}} \cdot u_{dyn}
\]

We set $P_{static}=100$ W and $P_{max}=400$ W, consistent with NVIDIA A100 specifications \cite{nvidia_nvidia_nodate}, with the upper and lower bound of utilization $u_{min}=0.1$ and $u_{max}=0.9$ respectively. Assuming each NVIDIA A100 GPU contains 15{,}000 cores, VM-level power consumption is estimated based on core-hour usage. As shown in Fig.~\ref{fig:power_and_qos} (a), the power consumption of VMs ranges from $4\times10^{-4}$ W to 52.67 W on a logarithmic scale.

\subsection{VM rescheduling procedure}
\label{tab:rescheduling}

Each arm represents a data center with a sequence of VM jobs. At state \(s\), let \(B_s\) denote the default batch and \(W_s\) a short lookahead window. Under activation, a batch \(C_s \subset W_s\) of the same size as \(B_s\) is selected to minimize power consumption. Let \(P_j\), \(Q_j\), and \(I_j\in\{0,1\}\) denote the power cost, QoS delay cost, and interaction indicator of job \(j\), respectively. The default and selected power consumptions are
\[
P_{\mathrm{def}}(s)
=
\sum_{j\in B_s} P_j,
\qquad
P_{\mathrm{chosen}}(s)
=
\sum_{j\in C_s} P_j.
\]
The delayed jobs are defined as \(D_s = B_s \cap [C_s]^c\), with QoS penalty
\[
C_{\mathrm{delay}}(s)
=
\sum_{j\in D_s} I_j Q_j.
\]
The active reward for arm \(i\) in state \(s\) is
\[
\begin{aligned}
&r_i(s)
=
\max\Big\{
0,\;
\lambda_{\mathrm{LMP}}
\big[
P_{\mathrm{def}}(s)
-
P_{\mathrm{sel}}(s)
\big] \\
&-
\lambda_{\mathrm{delay}}
C_{\mathrm{delay}}(s)
\Big\}
\end{aligned}
\]
where \(\lambda_{\mathrm{LMP}}= \$0.03/\mathrm{kWh}\) denotes the LMP and \(\lambda_{\mathrm{delay}}\) is the delay-cost multiplier. Passive arms receive zero reward, while active arms capture the electricity cost savings from VM rescheduling, but are penalized by QoS degradation due to delays in interactive jobs.

\end{document}